\documentclass[useAMS,usenatbib]{mn2e}

\usepackage{times}

\input{epsf.tex}

\title[Trapping of stars by a supercluster]
{On the trapping of stars by a newborn stellar supercluster}
\author[D. D. Carpintero and J. C. Muzzio]
{D. D. Carpintero$^{1,2}$\thanks{E-mail: ddc@fcaglp.unlp.edu.ar (DDC); 
jcmuzzio@fcaglp.unlp.edu.ar (JCM)} and 
J. C. Muzzio$^{1,2}$\footnotemark[1]\\
$^{1}$Facultad de Ciencias Astron\'omicas y Geof\'{\i}sicas de la Universidad
Nacional de La Plata, Argentina\\
$^{2}$Instituto de Astrof\'{\i}sica de La Plata, CONICET La Plata -- UNLP, 
Argentina}

\begin{document}

\date{Accepted. Received.}

\pagerange{\pageref{firstpage}--\pageref{lastpage}} \pubyear{}

\maketitle

\label{firstpage}

\begin{abstract}
Numerical experiments conducted by \citet{fke06} suggest that a supercluster may
capture up to about 40 per cent of its mass from the galaxy where it belongs.
Nevertheless, in those experiments the cluster was created making appear its
mass out of nothing, rather than from mass already present in the galaxy. Here
we use a thought experiment, plus a few simple computations, to show that the
difference between the dynamical effects of these two scenarios (i.e., mass
creation vs. mass concentration) is actually very important. We also present the
results of new numerical experiments, simulating the formation of the cluster
through mass concentration, that show that trapping depends critically on the
process of cluster formation and that the amounts of gained mass are
substantially smaller than those obtained from mass creation.
\end{abstract}

\begin{keywords}
methods: N--body simulations -- galaxies: dwarf -- galaxies: star clusters
\end{keywords}

\section{Introduction}

\citet*[][FKE06 hereafter]{fke06} have recently proposed the capture of old
stars by massive stellar superclusters during their formation process as a
possible explanation for the different age and metallicity populations found in
some clusters (e.g., $\omega$ Centauri). They used numerical experiments to show
that up to about 40 per cent of the initial mass of the cluster can be gained
from stars of the galaxy where the cluster belongs and they even suggest that
the captured mass might exceed the mass of the cluster in some cases.

One problem with the numerical simulations of FKE06 is that, since essentially
all the captures take place during the formation of the cluster, it is obvious
that the formation process itself should strongly affect the dynamics of capture
and, therefore, it is crucial to use an adequate model of the formation process
in order to get reasonable estimates of the gained mass. Nevertheless, in FKE06
the clusters are created as Plummer models whose masses increase linearly from
zero to their final values. In other words, the mass of the cluster is {\it
created} rather than, as it should, taken from mass already present in the
galaxy. Although they acknowledged that problem, FKE06 argued that, since the
cluster is much less massive than the galaxy, the adjustment of the galaxy
potential due to the cluster formation is a tiny effect, which is true, but of
little relevance to the process of capture. Besides, they perform a test using a
Plummer model of constant mass that starts with a large scalelength which is
subsequently reduced (i.e., simulating the collapse that forms the cluster),
obtaining almost exactly the same result as with the Plummer model with variable
mass. Although this outstanding coincidence seems to give strong support to the
results of FKE06 we will explain below that, in fact, it does not. 

Here we will show that, although the trapping effect invoqued by FKE06 indeed
exists, when the supercluster is created from mass already present in the galaxy
the amount of captured mass is substantially smaller than what they found. The
next section presents a thought experiment and some computed results to show
that the difference between aggregating matter already present and creating it
is absolutely crucial for the result of the capture process and, besides, we
explain why the test done by FKE06 does not avoid the problem of creating matter
from nothing. The third section describes our own numerical experiments, using
the FKE06 scenario and our own, and their results are presented in the fourth
section. The fifth and final section discusses our results.

\section{The dynamical difference between creating and concentrating mass}
\label{gedanken}

Let us consider a spherical galaxy with a cluster being formed at its centre, so
that we can apply Newton's theorems for spherical systems \citep[see,
e.g.,][]{bt08}, and let us further assume that, except for the mass related to
the cluster formation, the rest of the mass of the galaxy keeps its original
distribution. If, following FKE06, we start with a zero mass cluster and
increase its mass up to a final value, {\it all} the masses of the system will
experience an additional central force of an amount depending on the distance of
the mass to the centre of the system. If, instead, we mimic the cluster
formation selecting as the primordial cloud a sphere centered at the centre of
the stellar system, with a radius smaller than that of the system, and we take
from every spherical shell of that sphere a certain fraction of its mass and
move it to the centre of the system to form there the cluster, the result is
very different: 1) Any mass at a radius that places it outside the primordial
cloud will experience no extra force, because the mass within that radius will
be the same; 2) The masses within the radius of the primordial cloud will
experience new radial forces that will be very small near the border of the
cloud and will increase as we consider masses closer to the centre. Notice that,
as the largest differences between the two cases correspond to the largest radii,
they also involve the largest volumes within the galaxy. 

Of course, the parameter relevant to the capture process is not the force but
the potential: a star will be captured by the cluster if, after the cluster
formation, the potential at the location of the star is reduced by an amount
larger than one half the squared velocity of the star; that is, the quantity we
should be interested in is the variation of the potential due to the formation
of the cluster. We used the force in the previous discussion because, while in a
spherically symmetric case the force at a certain radius depends only on the
mass within that radius, the potential depends also on the distribution of mass
outside that radius \citep[see, e.g.,][]{bt08} and that would have complicated
the discussion. Nevertheless, if we supplement our thought experiment with a few
simple computations, we can use the potential rather than the force for our
analysis. Let us consider again a spherical galaxy with a cluster being formed
at its center ---either by creating or concentrating mass--- and let us assume
that, except for the matter used to form the cluster by concentration, the
density distribution of the galaxy is not altered by the formation of the
cluster. If the mass of the cluster is created, the difference in the potential
at a certain point of  the galaxy, before and after the creation of the cluster,
will be independent of that density distribution. Instead, when the cluster is
created taking matter from the primordial cloud centered at the center of the
system, the potential will not change outside that cloud and its change within
the cloud will depend only on the density distribution within the cloud.

\begin{figure}
\epsfxsize=240pt\epsfbox{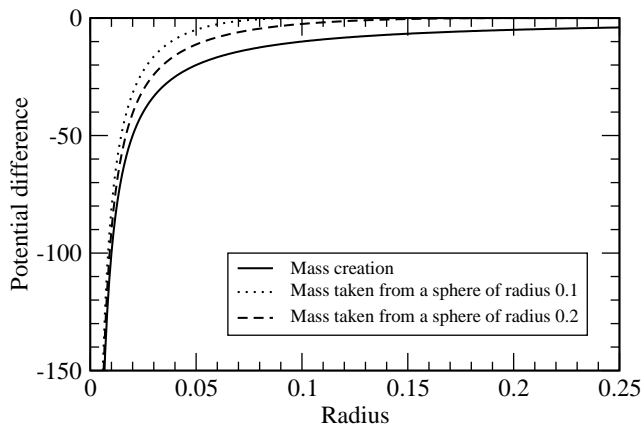}
\caption{The potential difference vs. radius when the cluster is formed by
either creating mass or concentrating mass already present within a primordial
cloud of density inversely proportional to radius for two different radii of the
cloud.} 
\label{fig01}
\end{figure}

A simple numerical example will illustrate this. Let us consider the cluster as
a point mass, $M_{\rm cls}=1$, and let us take the gravitational constant as
$G=1$. We took the density distribution of the  primordial cloud as $\rho (r) =
C r^{-\gamma}$, where $r$ is the radius, and $C$ and $\gamma$ are parameters of
the distribution (we recall that the density  distribution beyond the cloud
radius, $r_{\rm cld}$, is irrelevant for this computation); two different radii
were chosen for the primordial cloud, i.e., $r_{\rm cld}=0.1$ and $r_{\rm
cld}=0.2$. To form the cluster by concentration, we simply reduced $C$ by the
amount needed to take from the primordial cloud a mass equal to $M_{\rm cls}$
and we placed that mass at the centre of the galaxy. Fig. \ref{fig01} presents
our results: the full line gives the change in the potential for the case of
mass creation and the dotted and dashed lines give the same for the case of mass
concentration, respectively for $r_{\rm cld}=0.1$ and $r_{\rm cld}=0.2$; here,
we adopted $\gamma = 1$, but the result is very similar with $\gamma = 0$. Now,
if we draw a horizontal line at an ordinate equal to $-0.5 v^{2}$ where $v$ is
the star's velocity, the captured stars will be those at radii such that the
full line (in case of mass creation), or the dotted or dashed lines (in case of
mass concentration), lie below that horizontal line. Thus: 1) There will be
captured stars beyond $r=0.1$, or $r=0.2$, in the case of mass creation, but not
in the case of mass concentration; 2) Within $r=0.1$, or $r=0.2$, there will
always be more captures in the case of mass creation, and the difference with
the case of mass concentration will diminish as we go to smaller radii, becoming
zero only at the centre of the system; 3) The smaller the radius of the
primordial cloud, the larger is the difference between the mass creation and
mass concentration scenarios. As indicated previously, the difference between
the mass gain in both cases is smaller for smaller volumes, but we now see that
in those smaller volumes can be captured stars that move faster than those that
can be captured in larger volumes, so that there is some compensation of the
volume effect.

It is now evident that the dynamics of capture will be strongly affected by the
process of formation of the cluster and that creating matter leads to more
captures than concentrating it. Nevertheless, our thought experiment and simple
computations do not allow us to go beyond this qualitative conclusion and, to
reach quantitative results, we need to resort to numerical experiments. However,
before turning to them, let us analyze why the check perfomed by FKE06
attempting to simulate mass concentration, rather than creation, offers no check
at all.

FKE06 adopt their set-up corresponding to (small disc, heavy supercluster, one
scalelength distance), they create a cluster with a mass of $10^7 M_{\sun}$ and
a scalelength equal to the disc scalelength (0.5 kpc) and, finally, they shrink
that scale distance to the one of the cluster (25 pc) on a time scale equal to
the crossing time of the cluster (3.7 My). The problem is that, again, they
create mass from nothing. If we assume that the shrinking process is fast enough
so that the stars of the galaxy change their positions very little during that
process, it is obvious that, in the end, they would have gained essentially the
same (negative) potential energy as if the cluster had been created
instantaneously with the final scalelength. From Fig. 1 of FKE06 we can estimate
the velocity dispersion at one disc scalength radius as about 40 km s$^{-1}$ for
the small disc, that is, it will take an average star about 10 My to traverse
the scale length of the original cloud which is an interval long enough,
compared to that of the scalelength change, to accept that the stars have not
moved much during the shrinking process. In other words, the coincidence of the
result of this model with the original one is exactly what one could have
expected, and it is no proof that the creation of matter to build the cluster
does not affect the amount of gained mass.

\section{Description of the numerical experiments}

In order to establish a quantitative proof of the abovementioned qualitative
discussion, we performed a series of numerical experiments. We first set up a
background galaxy in equilibrium composed of a disc and an analytic halo,
without a bulge. The disc is realized using $5\times 10^6$ particles laid down
according to the following distribution function:
\begin{eqnarray}
f_{\rm d} & \propto & \exp\left(-\frac{R}{R_{\rm d}}\right)
                      {\rm sech}^2\left(\frac{z}{z_{\rm 0}}\right)\times 
                      \nonumber\\
& & \exp\left[-\frac{1}{2}\left(\frac{v_R^2}{\sigma_R^2}+
    \frac{(v_\varphi-v_{\rm d})^2}{\sigma_\varphi^2}+
    \frac{v_z^2}{\sigma_z^2}\right)\right],
\end{eqnarray}
that is, isothermal in the vertical direction with scaleheight $z_0$
\citep{s42},   exponential in the radial direction with scalelength $R_{\rm d}$,
and axisymmetric. The velocities are Gaussian, with dispersions $\sigma_z$,
$\sigma_R$ and $\sigma_\varphi$ in each direction respectively, and a mean
acimutal velocity $v_{\rm d}(R)$. The parameters $R_{\rm d}$ and $z_0$ are input
parameters, as well as the total mass of the disc $M_{\rm d}$. From these, the
dispersions and $v_{\rm d}(R)$ were computed following the recipe of
\citet{b92}. 

The potential of the halo is given by
\begin{equation}
\Phi(r)=v_0^2 \ln\left(r^2+R_{\rm c}^2\right),
\end{equation}
that is, a spherical logarithmic potential with asymptotic circular velocity
$\sqrt{2}v_0$ and core radius $R_{\rm c}$. Both $v_0$ and $R_{\rm c}$ are input
parameters.

\begin{figure}
\epsfxsize=240pt\epsfbox{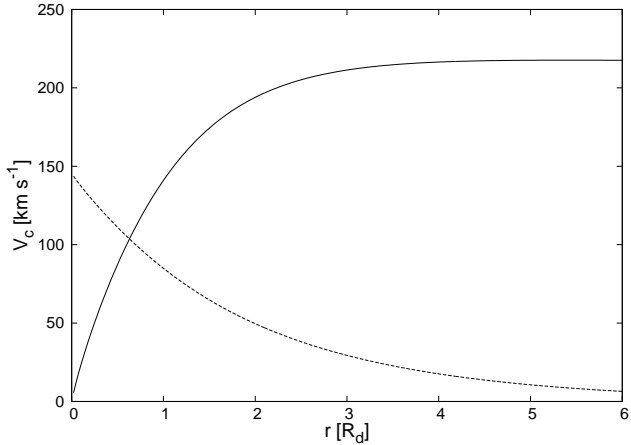}
\caption{Circular velocity (solid line) and three-dimensional velocity
dispersion (dashed line) of the galaxy, using $M_{\rm d}=10^{10} M_{\sun}$ and
$R_{\rm c}=2.5$ kpc.}
\label{fig02}
\end{figure}

We choose units such that the gravitational constant $G=1$, $R_{\rm c}=2.5$ and
$M_{\rm d}=10$. With this choice, we set $R_{\rm d}=1.5$, $z_0=0.25$, and
$v_0=2.287$. Using the equivalences $M_{\rm d}=10^{10} M_{\sun}$ and $R_{\rm
c}=2.5$ kpc, these units correspond to the high-mass galaxy of FKE06, although
our model differs somewhat from theirs in the velocity space, as can be seen
comparing Fig. \ref{fig02} with Figure 1 of FKE06.

Our cluster is built up in two different ways: a) By letting the potential of an
analytical Plummer sphere with scalelength $b_{\rm P}$ and total mass $M_{\rm
P}$ to grow from zero to its maximum strength. The growing up is achieved by
varying $M_{\rm P}$ linearly with time during an interval $t_{\rm P}$ equal to
the crossing time of the final Plummer sphere. The centre of the potential is
put in a circular orbit of radius $R_{\rm P}$. b) By letting a fraction of the
particles inside a sphere of total mass $M_{\rm s}$ and radius $r_{\rm s}$ to
move as if, besides their original velocities, they were in free fall with
respect to the centre of the sphere. The particles are randomly chosen among
those inside the radius $r_{\rm s}$. The total mass of the falling particles is
$M_{\rm ff}$, the mass of the future cluster. This in turn determines the radius
$r_{\rm s}$ as that which is required for $M_{\rm ff}$ to be the desired
fraction. The centre of the sphere is put in a circular orbit of radius $R_{\rm
ff}$. The free fall is achieved by adding the potential of a homogeneous sphere
chosen so that the free fall time is a desired value $t_{\rm ff}$. This
potential is not seen by the rest of the particles of the galaxy. The falling
particles are kept under the influence of the added potential until they reach a
small fiducial radius $b_{\rm ff}$ with respect  to the centre of the sphere,
from which point they are given the velocity of that centre and freed from the
falling. After that, if any of those particles leaves the sphere of radius
$b_{\rm ff}$, it is forced to fall again. The first approach (Plummer sphere)
implies assuming that the mass of the cluster is taken from outside the galaxy,
whereas in the second one (free fall sphere) the mass of the cluster is taken
from the galaxy itself.

\begin{table*}
 \caption{Parameters of the models}
 \label{tabla1}
 \begin{tabular}{@{}lccccccc}
  \hline
Model & $b_{\rm P}$,$b_{\rm ff}$  & 
${M_{\rm P}\over M_{\rm d}}$,${M_{\rm ff}\over M_{\rm d}}$ & 
$t_{\rm P}$,$t_{\rm ff}$  & $R_{\rm P}$,$R_{\rm ff}$  & 
${r_{\rm s}\over R_{\rm d}}$ & ${M_{\rm ff}\over M_{\rm s}}$ & 
${M_{\rm a}\over M_{\rm P}}$,${M_{\rm a}\over M_{\rm ff}}$ \\
  \hline
P1   & 0.025         & 0.002  & 0.175 & 1.5 & --   & --   & 0.099 \\
P1b  & 0.015         & 0.002  & 0.081 & 1.5 & --   & --   & 0.101 \\
C1   & 0.015         & 0.002  & 0.175 & 1.5 & 0.54 & 0.10 & 0.039 \\
P2   & 0.025         & 0.002  & 0.175 & 3.0 & --   & --   & 0.149 \\
C2   & 0.015         & 0.002  & 0.175 & 3.0 & 0.56 & 0.10 & 0.032 \\
C2b  & 0.015         & 0.002  & 0.175 & 3.0 & 0.77 & 0.05 & 0.042 \\
C2c  & 0.015         & 0.002  & 0.175 & 3.0 & 0.41 & 0.20 & 0.027 \\
P3   & 0.025         & 0.001  & 0.250 & 1.5 & --   & --   & 0.055 \\
C3   & 0.015         & 0.001  & 0.250 & 1.5 & 0.54 & 0.05 & 0.013 \\
P4   & 0.025         & 0.003  & 0.143 & 1.5 & --   & --   & 0.140 \\
C4   & 0.015         & 0.003  & 0.143 & 1.5 & 0.54 & 0.15 & 0.029 \\
P5   & 0.025         & 0.010  & 0.078 & 1.5 & --   & --   & 0.295 \\
C5   & 0.015         & 0.010  & 0.078 & 1.5 & 1.13 & 0.10 & 0.226 \\
C5b  & 0.015         & 0.010  & 0.078 & 1.5 & 0.81 & 0.20 & 0.198 \\
C5c  & 0.015         & 0.010  & 0.078 & 1.5 & 0.54 & 0.50 & 0.119 \\
S1   & 1.5 to 0.025  & 0.002  & 0.175 & 1.5 & --   & --   & 0.099 \\
S1b  & 0.25 to 0.025 & 0.002  & 0.175 & 1.5 & --   & --   & 0.098 \\
  \hline
 \end{tabular}
\end{table*}

Table \ref{tabla1} shows the parameters used for the different cluster models in
our experiments. Model names starting with P refer to experiments in which the
cluster is simulated through a Plummer sphere; names starting with C indicate
simulations in which the mass of the cluster is concentrated from the
environment, that is, the free fall generated clusters (although these models
are not free fall experiments in a strict sense, we will still call them free
fall models for simplicity). Model P1 is our basic model: the growing time
corresponds to the crossing time of the Plummer sphere, the radius of the
circular orbit equals the scalelength of the disc, and the mass of the cluster
is 1/500 of the mass of the disc. Model P1b is like model P1 but has a Plummer's
scalelength equal to the radius of the free fall final sphere. This is to verify
whether the difference between $b_{\rm P}$ and $b_{\rm ff}$ is affecting the
comparison between the Plummer and the free fall models. Model C1 has the same
mass as model P1, and the same circular orbit. The free fall time was chosen
equal to that of Model P1, and the free fall radius $b_{\rm ff}$ was chosen
1/100 of the scalelength of the disc, comparable to the scalelength of model P1.
The falling mass corresponds to a 10 per cent of the mass inside radius $r_{\rm
s}$. Models P2 and C2 are the same as P1 and C1, respectively, but the radius of
the circular orbit is doubled, in order to probe a different ambient for the
cluster. Models C2b and C2c are identical to model C2, but the falling masses
correspond to a 5 per cent and a 20 per cent of the mass inside the sphere of
radius $r_{\rm s}$, respectively. These models, which vary only the radius of
the sphere from which the mass to be concentrated is taken, allow a verification
of what was said in Section \ref{gedanken} with respect to changing the size of
the cloud. Models P3 and C3 are also the same as P1 and C1, but the respective
clusters have half the mass, and, correspondingly, a larger crossing/free fall
time. Models P4 and C4 have 1.5 times the mass of models P1 and C1,
respectively, and a corresponding shorter crossing/free fall time. These four
last models were run in order to assess how much the results are affected when
the mass of the cluster is changed. Model P5 corresponds to a Plummer sphere
that grows to a whole 1 per cent of the mass of the disc. Model C5 is the
corresponding free fall experiment, where the mass of the cluster is 10 per cent
of the mass inside the sphere of radius $r_{\rm s}$. Models C5b and C5c are
identical to model C5, but the falling masses correspond to a 20 per cent and a
50 per cent of the mass inside the sphere of radius $r_{\rm s}$, respectively.
Model S1 corresponds to a Plummer sphere that is born with all its mass, but
with an initial scalelength $b_{\rm P,0}$ equal to the scalelength of the disc,
$R_{\rm d}$. This scalelength is shrunk according to
\begin{equation}
b_{\rm P}(t)=(b_{\rm P,0}-b_{\rm P,f}){1-\exp(t-t_{\rm P})\over 1-\exp(-t_{\rm
P})}+b_{\rm P,f}\quad 0\le t\le t_{\rm P},
\end{equation}
where $b_{\rm P,f}$ is the final value of the scalelength, after a time $t_{\rm
P}$ equal to the crossing time of the final Plummer sphere. This model
corresponds to a cluster similar to that of the last numerical experiment of
FKE06 (by the way, there is probably an error in their Equation (3), since at
$t=0$ the Plummer radius is not the initial radius). Model S1b is similar to
model S1 but with the initial Plummer radius reduced to a sixth, in order to
probe whether the size of the initial radius has any influence in the capture of
mass during the shrinking stage. 

The experiments were run until $t=3$, corresponding to almost one period of the
cluster when put in a circular orbit of radius $R_{\rm d}$. The code used was a
{\tt FORTRAN77$+$MPI} version of the paralellized tree code of \citet{vc00}. It
ran in a cluster of twenty-four 1.86 GHz processors; each experiment took
approximately $10.5\times24$ hours of CPU time. 

In order to assess which particles were added to a cluster when modelled as a
Plummer sphere, following FKE06, we computed the energy of the particles with
respect to the sphere, as well as the tidal radius $r_{\rm t}$ of the latter. We
then considered as acquired by  the cluster those particles with both negative
energy and position inside $r_{\rm t}$. In order to determine the value of
$r_{\rm t}$ for each experiment, we  followed the working out of \citet[][\S
8.3.1]{bt08}, but replacing the acceleration of a point mass galaxy by that of
our disk plus halo system, and the acceleration of a point mass satellite by
that of our cluster. The resulting equation is:
\begin{eqnarray}
{GM_{\rm h}(R_{\rm P}-x)\over (R_{\rm P}-x)^2} 
+{V_{\rm cd}^2(R_{\rm P}-x)\over R_{\rm P}-x}
-{GM_{\rm C}(x)\over x^2}-{}&&\nonumber\\
GM_{\rm h}(R_{\rm P}){(R_{\rm P}-x)\over R_{\rm P}^3}
-V_{\rm cd}^2(R_{\rm P}){R_{\rm P}-x\over R_{\rm P}^2}=0,&&
\label{rmarea}
\end{eqnarray}
where $0<x<R_{\rm P}$, $M_{\rm h}(r)$ is the mass of the halo inside distance
$r$ of its center, $M_{\rm C}(r)$ is the mass of the cluster inside distance $r$
of its center, and $V_{\rm cd}^2(r)$ is the squared circular velocity of the
disk at a distance $r$ of its center given by \citep[see, e.g.][]{bt08}
\begin{equation}
V_{\rm cd}^2(r)=4\pi G \Sigma_{\rm d} R_{\rm d} y^2 \left[ I_0(y)K_0(y)
-I_1(y)K_1(y)\right],
\end{equation}
where $\Sigma_{\rm d}$ is the surface density of the disc, $y\equiv r/(2R_{\rm
d})$  and $I_0$, $I_1$, $K_0$ and $K_1$ are modified Bessel functions. The value
of $x$ that satisfies Eq. (\ref{rmarea}) is our tidal radius $r_{\rm t}$.

We also used the tidal radius as one of the criteria to define membership in the
free fall models. In these cases, however, we have replaced $M_{\rm C}(x)$ in
Eq. (\ref{rmarea}) by $M_{\rm ff}$, that is, the total mass of the cluster. This
amounts to considering the cluster as a point mass, which is a good
approximation provided that the free fall has already finished and that its
radius $b_{\rm ff}$ is smaller than  the computed $r_{\rm t}$ ---that was the
case in all the experiments. The other criterion, negative energy, was computed
by first finding which particles were geometric neighbours of the center of the
free fall with the aid of a friend-of-friend algorithm, taking 0.70 of the mean
interparticle distance of the 90 per cent most bounded disc particles as the
fiducial maximum neighbour distance, which sufficed to neatly isolate the
cluster from its surroundings. We then computed the energy of these particles
with respect to the set, and discarded those with positive energy and/or outside
the tidal radius. This step --computation of the energy and discarding-- was
repeated with the remaining particles until only particles inside the tidal
radius and with negative energy were left; these particles were considered the
members of the cluster. Also, during the free fall, the list of members was
considered empty, since the tidal radius along that period is not well defined.

\section{Results}

\begin{figure}
\epsfxsize=240pt\epsfbox{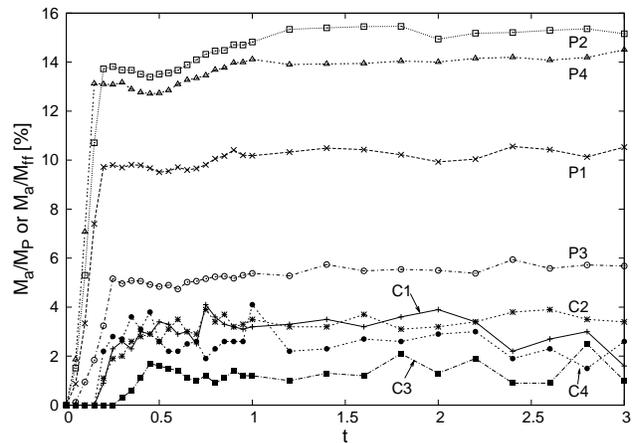}
\caption{Percentage of mass acquired by the cluster in models P1 through P4 and
C1 through C4, as a function of time.}
\label{fig03}
\end{figure}

Fig. \ref{fig03} shows the percentage of trapped particles of the basic models
P1 and C1, when the orbit of the cluster is at $2R_{\rm d}$ (models P2 and C2),
and when the mass of the cluster is half and one and a half that of models P1
and C1 (models P3, C3, P4 and C4, respectively). It is clearly seen that the
analytical Plummer model traps more mass than the free fall in all the cases,
that is, the trapped mass depends on whether the mass of the cluster is taken
from outside the galaxy or from the galaxy itself. Model P3 (the less masive
Plummer model) is the only one that captures a mass comparable with the free
fall models (but substantially larger than that captured by its equivalent model
C3). In all cases, the fluctuations in the captured mass after the cluster
finished its growth are due to particles close to the tidal radius and with
energies close to zero, therefore oscilating between trapped and non trapped
stages.

Table \ref{tabla1} shows in its last column the mass $M_{\rm a}$ acquired by the
cluster in each experiment, as a fraction of the mass of the cluster $M_{\rm P}$
or $M_{\rm ff}$. These data were taken at a representative time $t=2$, that is,
after about two thirds of an orbit of the cluster since its birth, a long enough
interval compared to the growing time of any of the models. Clearly, all
the free fall models captured considerably less mass than the corresponding
Plummer models. We can see that there is no significant difference between model
P1 and model P1b; therefore, the accreted mass does not depend significantly on
the details of the final scalelength of the Plummer sphere. We can also see that
model C2b traps more mass than model C2: as we had anticipated, it is the
expected behaviour when a larger initial radius $r_{\rm s}$ is used. Model C2c,
on the other hand, having a smaller $r_{\rm s}$ than model C2, acquires a little
less mass than the latter.

The heavy models P5, C5, C5b and C5c also show the same trends. Although in the
case of model C5 the trapped mass is of the same order of that in the
corresponding Plummer model, the other two acquired much less mass. These models
clearly show that the percentage of trapped mass depends on how the mass of the
cluster is gathered from the galaxy: the smaller the region of ambient mass that
is used to build the cluster, the more additional mass is acquired after the
cluster is formed.

Finally, the trapped mass of the experiments S1 and S1b, as expected, is almost
the same as in model P1, that is, the shrinking of the scalelength of the
Plummer sphere has no effect whatsoever on the accumulated mass, as could be
expected from our discussion of Section \ref{gedanken}. Besides, since the
crossing time of the original cloud is reduced by to one--sixth when going from
model S1 to model S1b (from 1.54 units of time to 0.25), whereas the shrinking
time is held constant (equal to the  crossing time of the final configuration,
0.175 units of time), a typical galactic star can cross almost the entire radius
of the cloud in the time that that cloud reduces its size. Therefore, the
assumption of instant collapse, adopted in the discussion of Section
\ref{gedanken}, is not critical.

\section{Conclusions}

Our results confirm the finding of FKE06 that, during the formation of a
supercluster in a dwarf galaxy, some mass can be additionally gained from
trapped disc stars and that the capture process essentially ends with the
formation of the cluster, with virtually no gains afterwards; the exceedingly
small amount of captures by an already formed cluster has been also found by
\citet{mb07}.

Nevertheless, while FKE06 do not assign much importance to the process of
formation of the cluster and simply simulate it with a mass that grows linearly
with time from zero to its final value, we consider that the details of such
process are crucial for the trapping dynamics. We have shown in Section
\ref{gedanken} that, in particular, creating the mass of the cluster from
nothing originates forces and potentials fairly different from those that appear
when the cluster is formed concentrating matter already present in the galaxy
and that, as a result, less trapping should be expected from the latter
scenario.

We performed several numerical simulations similar to those of FKE06, where the
mass of the cluster is created out of nothing (our P models), together with
others that only differ from the former in that the cluster is formed
concentrating mass from the galaxy (our C models). In all cases, the mass gained
by the C models was smaller than that gained by the P models, the most extreme
examples being those of models C2c and C4 which gain only about one--fifth of the
mass gained by their equivalent models P2 and P4, respectively.

The difference in gain depends critically on the size of the primordial cloud
from which the C models get their cluster material: the smaller the primordial
cloud, the larger the difference in gained material. Since in our models the
mass taken from the cloud to build the cluster is uniformly distributed all over
the cloud, the size of the cloud correlates inversely with the fraction of mass
taken, i.e., the larger the fraction of mass the smaller the cloud. Most of our
models take that fraction between 0.05 and 0.20, i.e., one might assume that
that is the fraction of gas in the galaxy and that all the gas within a certain
region (our primordial cloud) collapses to form the cluster. As a result, the
less massive clusters are formed from smaller regions and for them the
differences between the mass creation and mass concentration scenarios are the
largest. On the other hand, our model runs into trouble for the most massive
superclusters. To create a supercluster with one--hundredth the mass of the
galaxy we need either to assume an implausibly high fraction of collapsing mass
of 0.50 (model C5c) or, for a more reasonable fraction of 0.10 (C5) or 0.20
(C5b), to accept that the mass comes from a primordial cloud of radius $1.13
R_{\rm d}$ ($\simeq 1.7$ kpc) or $0.81 R_{\rm d}$ ($\simeq 1.2$ kpc),
respectively. Now, clouds of such size should be suffering the effect of the
differential rotation and the tidal forces of the galaxy, making very unlikely
their collapse to form the supercluster. The formation of such a huge
supercluster probably proceeds by separate stages, with smaller clusters being
formed first and later coallescing to create the supercluster, so that the
amounts of trapping predicted for this case by the simple models of FKE06 and
ours should be regarded, at best, as very doubtful.

Cluster formation is certainly a very complex process with effects ignored by
the models of FKE06 and ours, such as gas dynamics and magnetic fields playing a
significant role \citep[see, e.g.][]{sp04}, and supercluster formation is
probably even more complex. It is clearly an understatement to say that our
models are only a very crude representation of the dynamics of this process, but
our point is precisely that, since the trapping takes place during the cluster
formation, it is vital to take into account the details of that process to
correctly evaluate the amount of matter trapped. Crude as they are, our models
have over those of FKE06 the big advantage that they use mass already present in
the galaxy in a way that is undoubtedly far from how real clusters are formed,
but which is certainly closer to reality than creating mass from nothing.
Moreover, the results of our models confirm what a simple reasoning suggests,
i.e., the amount of matter trapped in our somewhat more realistic scenario of
mass concentration is substantially smaller than that which results from
creating the mass of the cluster out of nothing.

\section*{Acknowledgments}

The technical assistance of Ruben E. Mart\'inez and H\'ector R. Viturro and
helpful discussions with G. L. Bosch on cluster formation are gratefully
acknowledged. This work was done with the support of grants from the Universidad
Nacional de La Plata, the Agencia Nacional de Promoci\'on Cient\'ifica y
Tecnol\'ogica and the Consejo Nacional de Investigaciones Cient\'ificas y
T\'ecnicas de la Rep\'ublica Argentina (CONICET). In particular, the cluster of
twenty four processors (Athena) used for this work was built and configured by
H. R. Viturro, and paid with those grants.

\bsp

\label{lastpage}

\end{document}